# NLOS UV Channel Modeling Using Numerical Integration and an Approximate Closed-Form Path Loss Model

Ankit Gupta , Mohammad Noshad  and Maïté Brandt-Pearce

Charles L. Brown Department of Electrical and Computer Engineering, University of Virginia, Charlottesville, VA 22904

## ABSTRACT

In this paper we propose a simulation method using numerical integration, and develop a closed-form link loss model for physical layer channel characterization for non-line of sight (NLOS) ultraviolet (UV) communication systems. The impulse response of the channel is calculated by assuming both uniform and Gaussian profiles for transmitted beams and different geometries. The results are compared with previously published results. The accuracy of the integration approach is compared to the Monte Carlo simulation. Then the path loss using the simulation method and the suggested closed-form expression are presented for different link geometries. The accuracies are evaluated and compared to the results obtained using other methods.

## 1. INTRODUCTION

Ultraviolet communications has attracted significant interests for application in non-line of sight (NLOS) free space communication in recent years. Low background noise and high atmospheric scattering make the UV-C band (200-280 nm) a good option for NLOS communications. Attenuation and inter-symbol interference (ISI) are the limiting factors in NLOS UV links and an accurate evaluation of the performance in these systems requires a precise model for both the path loss and impulse response.[1]

Many techniques have been presented to accurately model the NLOS UV channel. A simulation approach that uses a Monte Carlo method for multiple scattering interactions is presented by Ding et al.[2] and by Drost et al.[3] An analytical approximation can be remarkably helpful for getting a fast initial scope of the link performance. Luettgen, Shapiro, and Reilly[4] present an analytical model in integral form for the impulse response by considering only the single scatter propagation. Xu[5] analyzes the performance of NLOS UV links using a single scattering approximation. An approximate closed form path loss for NLOS UV links with small transmitted beam-widths and small fields of view (FOV) is calculated.[5] The common volume between the transmitted beam and receiver FOV is approximated by a frustum and, in this way, a closed form expression for path loss is obtained.[6] Recently, another approximate link loss is calculated for noncoplanar geometries.[7]

In this paper we propose a simulation method using numerical integration. In this approach, the space is divided into small cubic volumes and the received energy from each volume and its arrival time are calculated. The impulse response is calculated based on these time samples using a kernel function. The path loss is calculated as the ratio of the transmitted energy over the total received energy from all small volumes. The numerical integration technique provides advantages of scalability and flexibility while also accounting for shadowing due to obstructive objects in the system. This is compared to the analytical approach,[4] which, although faster, cannot incorporate shadowing effects in the system model. A comparison is also drawn with the Monte Carlo approach as described by Drost et al.[3] While the Monte Carlo approach can incorporate shadowing effects, for the same accuracy, the proposed numerical integration approach is computationally more efficient. For the same number of trials or cubes, the algorithm complexity is lower than the Monte Carlo approach.

Pointing requirements for UV NLOS systems are stringent for narrow beams since misdirected transmitter-receiver pairs result in smaller overlap volumes and hence, increase path loss significantly. Spatial multiplexing

E-mail: ag8ar@virginia.edu, mn2ne@virginia.edu, mb-p@virginia.edu

provides a solution to this problem. Multiple narrow transmit beams can be used to ensure overlap with the receiver beam even in cases of high transmitter-receiver azimuth skew. The proposed method can analyze multiple-input multiple-output (MIMO) systems by incorporating scalability for multiple transmitter receiver pairs. The numerical integration approach is especially computationally efficient for these MIMO systems.

Additionally, an analytical approach for calculating the path loss is developed by dividing the receiver FOV into differential angles and integrating over the generated differential volumes. The previously presented closed-form path losses[5,6] calculate the scattered energy from only one point, and approximate the total received energy by multiplying it by the common volume between the transmitted beam and receiver FOV. The error in these approximations can be large for receivers with wide FOV. In our proposed approach we calculate the received energy for each differential angle inside the FOV and integrate over the FOV. Hence, even for receivers with wide FOV, the proposed analytical path estimate remains accurate.

The rest of the paper is organized as follows. In Section 2, the proposed simulation approach is explained and its accuracy is compared to the Monte Carlo method. The analytical model for the link loss is developed in Section 3. Results are presented in Section 4, and the accuracy of the proposed methods is compared with previously published results. Section 5 summarizes the approaches presented and offers conclusions.

## 2. SIMULATION APPROACH FOR IMPULSE RESPONSE

Following the framework of previous work,[6,8] we consider a somewhat idealized model of the communication channel. Suppose a UV transmitter directs radiation within an ideal circular cone with cone angle $\phi_1$, elevation angle $\theta_1$ and an azimuth angle of $\alpha_1$, as illustrated in Fig. 1. Light within this cone is scattered from natural atmospheric particles and other aerosols. The receiver whose detector has a finite field of view, or acceptance region, is modeled by an ideal cone with angle $\phi_2$, elevation angle $\theta_2$ and azimuth angle $\alpha_2$. To first-order, these two overlapping cones define a common scattering volume $V$ in the sky, and only energy scattered from within this volume is considered as captured by the receiver. Let $E_t$ denote transmitter energy, $\Omega_t$ the transmitter solid angle in steradian, and $A_r$ the receiver collecting aperture. $r$ denotes the transmitter ($T_x$) and receiver ($R_x$) baseline separation, and $r_1$ and $r_2$ the distances of the common volume to the $T_x$ and $R_x$, respectively. $\theta_s = \theta_1 + \theta_2$ is the angle between the forward direction of the transmitted waves and the receiver axis.

Assuming a single-scattering model, Luettgen, Shapiro, and Reilly[4] develop a propagation model that accounts for scattering and absorption as exponential attenuations with distance. To calculate the total energy received for all transmitter and receiver geometries, we divide the volume enclosing the transmitter and receiver pair into smaller cubical differential volumes ($\delta V$). For an isotropic transmitter and receiver pair, each of these cubical volumes contributes to a differential energy ($\delta E_r$) at the receiver as in Eq. [4, (3.4)].

To take into account the system geometry and beam parameters, both the transmitter and receiver are modeled using gain factors $G_t$ and $G_r$, respectively. The receiver is modeled with a Lambertian acceptance profile. Thus, for all angles within the field of view $\phi_2$ of the receiver, a receiver Lambertian gain $\delta G_r$ is observed as $\delta G_r = \cos(\xi)$, where $\xi$ is the angle made between the receiver axis and the vector from the center of each differential volume to the receiver. For the transmitter, we consider two cases: one modeling the transmitter beam with a Gaussian profile and the second modeling the transmitter energy as uniformly distributed over the transmitter solid cone angle. Each differential volume experiences an effective transmitter gain factor $\delta G_t$. The transverse gain profile of the transmitter beam as modeled with a Gaussian function is given by[9]

$$\delta G_t = \frac{1}{\pi(w_z^2)} e^{-2r_t^2/w_z^2}, \tag{1}$$

where the beam radius, $w_z$, is the set of points in the $1/e^2$ contour after the wave has propagated a distance $z$ along the beam axis, and $r_t$ is the transverse distance of the center of each differential volume from the central axis of the beam. The beam radius varies along the propagation direction $z$ according to $w_z = w_o(1 + z/z_r)^{1/2}$, where $z_r$ is the Rayleigh length and $w_o$ is the radius of the $1/e^2$ irradiance contour at the beam waist. Similarly, for a uniform beam, the gain of the transmitter is given by

$$\delta G_t = \begin{cases} 1/\Omega_t & \text{if } \delta V \quad V \\ 0 & \text{otherwise.} \end{cases}$$

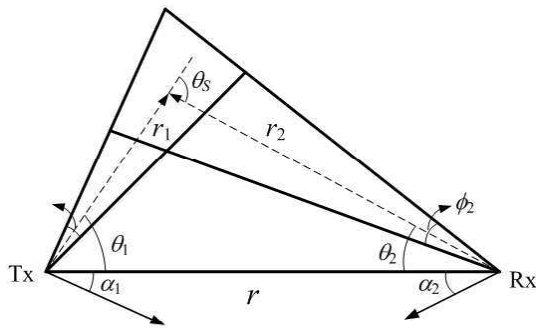

Figure 1. Geometry of the NLOS UV link.

A differential volume ($\delta V$) contributes differential energy (in Joules) at the receiver given by[4]

$$\delta E_r = \frac{k_s E_t P(\mu) A_r \delta V}{r_1^2 r_2^2} e^{-k_e(r_1+r_2)}. \tag{2}$$

This differential energy per solid angle received assumes an isotropic transmitter and receiver pair. The scattering coefficient, $k_s$, is a function of wavelength and the scattering angle $\theta_s$. The exponential extinction coefficient, $k_e$, is the sum of $k_s$ and the absorption coefficient, $k_a$. $k_s$ is given as $k_s = k_s^{Mie} + k_s^{Ray}$, where $k_s^{Mie}$ is the scattering coefficient due to Mie or aerosol scattering, and $k_s^{Ray}$ is the scattering coefficient due to Rayleigh or molecular scattering. $P(\mu)$ is the phase scattering function with $\mu = \cos(\theta_s)$. This is modeled as a weighted sum of the Rayleigh (molecular) and Mie (aerosol) scattering phase functions based on the corresponding scattering coefficients,[10]

$$P(\mu) = [P_{Ray}(\mu) + \frac{k_s^{Mie}}{k_s^{Ray}} P_{Mie}(\mu)]/(1 + \frac{k_s^{Mie}}{k_s^{Ray}}) \tag{3}$$

The Mie scattering phase function is given by a generalized Henyey-Greenstein function[10]

$$P_{Mie}(\mu) = \frac{1-g^2}{4\pi}[(1+g^2-2g\mu)^{-3/2} + f\frac{0.5(3\mu^2-1)}{(1+g^2)^{3/2}}] \tag{4}$$

where $g$ is the aerosol asymmetry parameter given by the mean cosine of the scattering angle, $f$ is aerosol hemispheric backscatter fraction. The Rayleigh scattering phase function is modeled as a generalized Rayleigh model[11]

$$P_{Ray}(\mu) = \frac{3}{4(1+2\gamma)}[(1+3\gamma) + (1-\gamma)\mu^2], \tag{5}$$

where $\gamma$ is defined by[12]

$$\gamma = \frac{\rho}{2-\rho}. \tag{6}$$

$\rho$ is the depolarization parameter.

For each $\delta V$, the receiver detects effective energy $\delta E_e$ Joules given by $\delta E_r \times \delta G_t \times \delta G_r$. Thus, once we calculate the received energy $\delta E_r$, which is fixed for a particular transmitter to receiver distance, it can be multiplied by the transmitter and receiver gains for different geometries to provide actual energy received from each $\delta V$. The total energy received is the total contribution of all such energies corresponding to each of the differential volumes integrated over the common scattering volume $V$:

$$E_r = \sum_V \delta E_r \times \delta G_t \times \delta G_r. \tag{7}$$

This received energy is then evaluated numerically by replacing the differential volumes with discrete cubes and summing their individual contributions. All differential geometry parameters used in this equation are calculated with respect to the center of each of the differential cubes.

The impulse response of the channel is also calculated using the same method. It is given by the expected energy received through the multipath channel with the appropriate arrival time delay. Using the numerical integration approach for path loss, the same differential volumes encompassing the $T_x$ and $R_x$ beams are used. For each volume, the energy received per unit time is noted versus the time delay for the path between the $T_x$ and $R_x$, $t = (r_1 + r_2)/c$ (where $c$ is the speed of light), passing through the center of the differential volume. This energy received is now a function of the time delay for each path corresponding to the differential volumes. The energy per unit time as a function of the energy arrival times is convolved with a kernel function to smooth the discrete differential contributions to the impulse response. To ensure sufficient accuracy in the curve and minimize contributions due to overlapping tails of each kernel functions, we use a narrow Gaussian kernel.

Using the numerical integration technique, we compute the total received energy for different $T_x$ and $R_x$ geometries varying by elevation, beam, azimuth angles and transmitter-receiver distance. This technique is computationally efficient and can provide extra flexibility by accounting for noncoplanar $T_x$-$R_x$ geometries as well as shadowing in the system due to objects obstructing the propagation path between the transmitter and receiver. This shadowing factor can be incorporated in the transmitter and receiver gain terms, and provides an added advantage over previous work.[4] The Monte Carlo approach as described by Drost et al.[3] can deal with both noncoplanar geometries and shadowing but requires more computations for a similar accuracy. We implement the numerical simulation using a composite midpoint rule for integrating over the large volume $V$ the energy $\delta E_r$ due to each of the differential volumes $\delta V$. The error of this algorithm is upper bounded by $O(N^{-2/3})$ and its complexity is given by $O(N)$, where $N$ is the total number of differential volumes making up the space encompassing the system. Typical values of $N$ that were used in the simulations were on the order of $10^6$. This method provides higher accuracy for a similar number of trials when compared to the Monte Carlo simulations.[3] For $P$ trials, the Monte Carlo simulations[3] are complexity bounded by $O(P)$ with the error order of convergence as $O(P^{-1/2})$. Thus, for the same computational overhead, better accuracy is achieved using the numerical integration approach.

An added advantage of the numerical integration approach is its flexibility in dealing with MIMO systems. Using this approach, the received energy matrix for each cube ($\delta E_r$) needs to be calculated only once, after which the receiver and transmitter gains ($\delta G_t, \delta G_r$) can be superimposed for any number of transmitters and receivers. This reduces the computational overhead significantly. Moreover, for larger distances, the whole system can be scaled by using the same number of cubes but now with a larger dimension. This makes the numerical integration method more convenient to use even for different distances.

## 3. ANALYTICAL APPROXIMATION FOR LINK LOSS

In this section, a closed form analytical approximation for the loss in the NLOS UV channel is presented.

The geometry of a NLOS UV channel is illustrated in Fig. 2. According to previous work,[6] the received energy from volume $dV$ is

$$dE_r \approx \frac{E_t k_s P(\mu) A_r dV \sin^4(\theta_1 + \theta) \exp[-k_e r \frac{\sin\theta_1 + \sin\theta}{\sin(\theta_1+\theta)}]}{2\pi r^4 \sin^2\theta_1 \sin^2\theta (1 - \cos\frac{\phi_1}{2})}, \tag{8}$$

where $E_t$ is the total transmitted energy and $\theta$ is the angle between the base line and the line that connects the center of the volume to the receiver, as shown in Fig. 2. According to Fig. 2, we divide the receiver FOV into differential angles. We assume that $\phi_1 < \phi_2$, which holds for practical NLOS UV links. In this way, the volume intersecting the transmitted beam and the differential angle $d\theta$ can be approximated as a elliptical-cylinder. The volume of this cylinder is equal to $A_s dx$, where $A_s$ is the area of its elliptical surface, and $dx$ is the thickness of the cylinder given by $dx = r_2 d\theta$. According to Fig. 3, $A_s$ is equal to $A'_s/\sin\theta_S$, where $\theta_S = \theta_1 + \theta$ and $A'_s$ is a surface perpendicular to the transmitter axis, and therefore a circle. Thus $A_s$ is equal to $r_1^2 \phi_1^2 \pi/(4\sin(\theta_1+\theta))$.

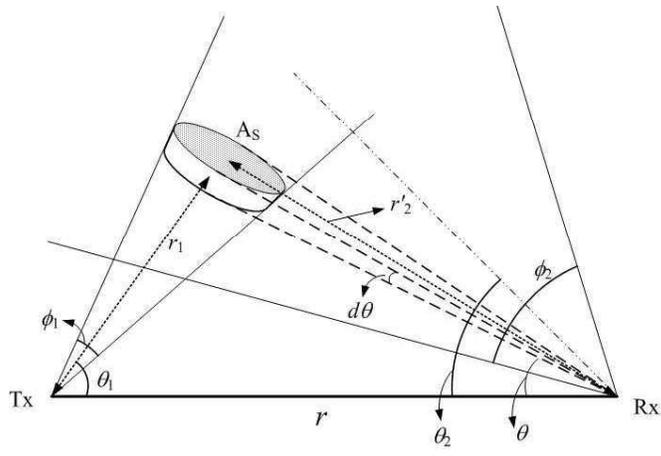

Figure 2. The intersection volume between the transmitted beam and differential FOV, $d\theta$.

Using these parameters, $dV$ becomes

$$dV = \frac{\pi}{4\sin(\theta_1+\theta)} r_1^2 r_2 \phi_1^2 d\theta. \tag{9}$$

Using (9), (8) simplifies to

$$dE_r = \frac{E_t k_s P(\mu) A_r \phi_1^2 d\theta \exp[-k_e r \frac{\sin\theta_1+\sin\theta}{\sin(\theta_1+\theta)}]}{8r\sin\theta_1(1-\cos\frac{\phi_1}{2})}. \tag{10}$$

The total received energy can be obtained by integrating over $dE_r$ as

$$E_r = \int dE_r = \int_{\theta_2-\frac{\phi_2}{2}}^{\theta_2+\frac{\phi_2}{2}} \frac{E_t k_s P(\mu) A_r \phi_1^2 \exp[-k_e r \frac{\sin\theta_1+\sin\theta}{\sin(\theta_1+\theta)}]}{8r\sin\theta_1(1-\cos\frac{\phi_1}{2})} d\theta, \tag{11}$$

where $\mu$ is assumed to be $\cos(\theta_1+\theta_2)$.

Defining the parameter $\tau$ as

$$\tau = \frac{1-\cos(\theta_1+\theta)}{\sin(\theta_1+\theta)}, \tag{12}$$

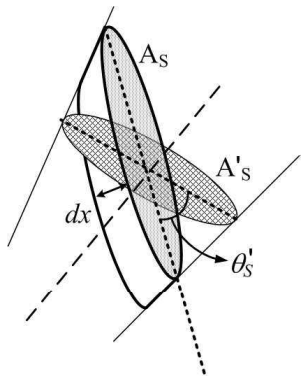

Figure 3. The elliptical-cylinder differential volume with thickness $dx$ and surface area $A$.

we have

$$E_r = X \int_{\tau_1}^{\tau_2} \frac{2}{1+\tau^2} e^{-B\tau} d\tau, \quad (13)$$

where

$$X = \frac{E_t k_s P(\mu) A_r \phi_1^2}{8r \sin\theta_1 (1 - \cos\frac{\phi_1}{2})} e^{-k_e r \cos\theta_1}, \quad (14)$$

$$\tau_1 = \frac{1 - \cos(\theta_2 - \frac{\phi_2}{2})}{\sin(\theta_2 - \frac{\phi_2}{2})}, \quad (15)$$

$$\tau_2 = \frac{1 - \cos(\theta_2 + \frac{\phi_2}{2})}{\sin(\theta_2 + \frac{\phi_2}{2})}, \quad (16)$$

and $B = k_e r \sin\theta_1$. In obtaining (13), we used

$$\sin(\theta_1 + \theta) = \frac{2}{\tau + 1/\tau}, \quad (17)$$

and

$$d\theta = \frac{\sin^2(\theta_1 + \theta)}{1 - \cos(\theta_1 + \theta)} d\tau = \frac{2}{\tau^2 + 1} d\tau. \quad (18)$$

Integrating over $\tau$ in (13) gives us

$$E_r = \left[ 2Xe^{-B\tau} \tan^{-1}\tau \right]_{\tau_1}^{\tau_2} + 2XB \int_{\tau_1}^{\tau_2} \tan^{-1}\tau \, e^{-B\tau} d\tau. \quad (19)$$

We approximate $\arctan\tau$ in the second term of (19) by

$$\tan^{-1}\tau = C(\tau_1, \tau_2)\tau + D(\tau_1, \tau_2), \quad (20)$$

over interval $[\tau_1, \tau_2]$, where $C(\tau_1, \tau_2) = \frac{\tan^{-1}\tau_2 - \tan^{-1}\tau_1}{\tau_2 - \tau_1}$ and $D(\tau_1, \tau_2) = \frac{\tau_2 \tan^{-1}\tau_1 - \tau_1 \tan^{-1}\tau_2}{\tau_2 - \tau_1}$. Substituting (20) in (19) gives us

$$E_r \approx 2C(\tau_1, \tau_2) \frac{X}{B} \left[ e^{-B\tau_1} - e^{-B\tau_2} \right]. \quad (21)$$

In (21) we approximate the received energy by integrating over the differential received energy in the receiver FOV, and therefore, by increasing the FOV, unlike the previously presented closed-form path loss approximations, the path loss presented in (21) stays close to the exact value.

## 4. NUMERICAL RESULTS

In this section numerical results for the impulse response and path loss are presented, and the results from different methods are compared.

Fig. 4 shows the impulse responses for $\theta_1 = 90°$, $\theta_2 = 90°$, $\phi_1 = 17°$, $\phi_2 = 30°$ and $r = 100$ m, using our numerical integration and the Luettgen-Shapiro-Reilly approach.[4] For the numerical integration method we have plotted two cases, one for the uniform transmitted beam and the other for the Gaussian beam. All results are obtained assuming only single scatter interactions. The impulse response curve for the Gaussian transmitter appears to rise earlier because of the volumes outside of the transmitter beam-width. The slight difference between the Luettgen-Shapiro-Reilly approach and the simulation for a uniform transmitted beam are because of the infinite common space between the transmitted beam and receiver FOV (for vertically pointed beams). In this case, since we only consider a finite number of volumes and do not account for the energy scattered from the upper space, a small error appears in the tail of the impulse response. This error can be reduced by using a larger space (more cubes) or modeling the space around the transmitter and receiver pair as cuboid with more height and correspondingly smaller width and distance. This makes the approximation more accurate by having more cubes cover the infinite vertical overlap. For sufficient accuracy, we cut off the infinite overlap vertically at $2 \times r_1 sin(\theta_1)$.

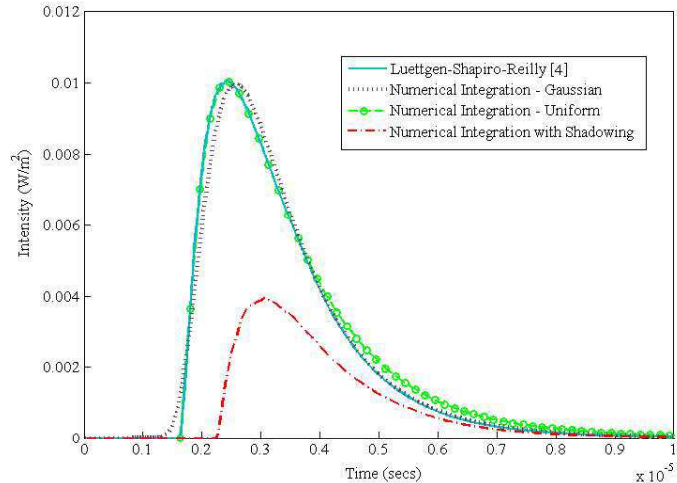

Figure 4. Impulse response for $\theta_1 = 90°$, $\theta_2 = 90°$, $\phi_1 = 17°$, $\phi_2 = 30°$, and $r = 100$ m using the Luettgen-Shapiro-Reilly approach[4] and the numerical integration method. Numerical integration is plotted for both uniform and Gaussian transmitted beams, and also for the case considering the shadowing effect.

Fig. 4 also shows the effects of shadowing on this configuration by displaying the impulse response curve obtained when a building of dimensions 30 m × 20 m × 150 m (length × width × height) is at distance of 5 m from the $T_x$. The length of this building is along the $T_x$-$R_x$ axis, the width is perpendicular to this axis on the ground plane, while the height is the vertical reach of the building. Thus this tall building shadow nearly halves the transmitter energy, which explains the lower area under the impulse response curve. The curve appears delayed since this building blocks shorter delay paths for the given configuration and the energy received is through the longer delay paths.

The impulse response for the geometry in which the receiver inclination angle, $\theta_2$, is deceased to 45°, is depicted in Fig. 5. Fig. 5 also shows shadowing effects on the impulse response when a shorter building (30 m × 20 m × 35 m) is at a distance of 40 m from the receiver. The impulse response for shadowing shows both attenuation and delay. Moreover, for this case the numerical integration results for the uniform transmitted beam matches well the Luettgen-Shapiro-Reilly plot. Note that unlike the previous geometry, the overlap volume between the transmitted beam and the receiver FOV is limited. The response time for this case is also shorter than the case of Fig. 4. Although our technique is flexible, no analytical methods exist in the literature for noncoplanar geometries or shadowed links, making such comparisons impossible.

The path loss of the NLOS UV link is depicted in Fig. 6 for $\theta_1 = 80°$, $\theta_2 = 60°$, $\phi_1 = 10°$ and $\phi_2 = 30°$. The scattering and extinction coefficients are considered to be $k_s^{Ray} = 0.24$ km$^{-1}$, $k_s^{Mie} = 0.25$ km$^{-1}$ and $k_a = 0.9$ km$^{-1}$. The path loss is plotted for measurement from Xu et al.,[6] Luettgen-Shapiro-Reilly method,[4] our proposed numerical integration method from Section 2, closed form expression from (21), and the analytical approximation calculated by Xu et al.[6] As can be seen the path loss using the numerical integration exactly matches the Luettgen-Shapiro-Reilly method plot. Also the proposed closed-form approximation is closer to the Luettgen-Shapiro-Reilly and numerical integration plots than the approximate expression presented by Xu et al.[6] The difference between the measurement results and the other results in all figures can be because of a difference in the modeling a real situation, or the effect of some un-modeled parameters.

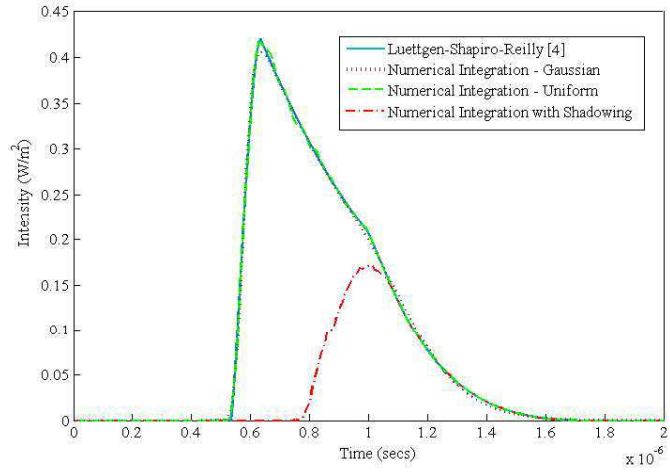

Figure 5. Impulse response for $\theta_1 = 90°$, $\theta_2 = 45°$, $\phi_1 = 17°$, $\phi_2 = 30°$, and $r = 100$ m using the Luettgen-Shapiro-Reilly approach[4] and the numerical integration method. Numerical integration is plotted for both uniform and Gaussian transmitted beams, and also for the case considering the shadowing effect.

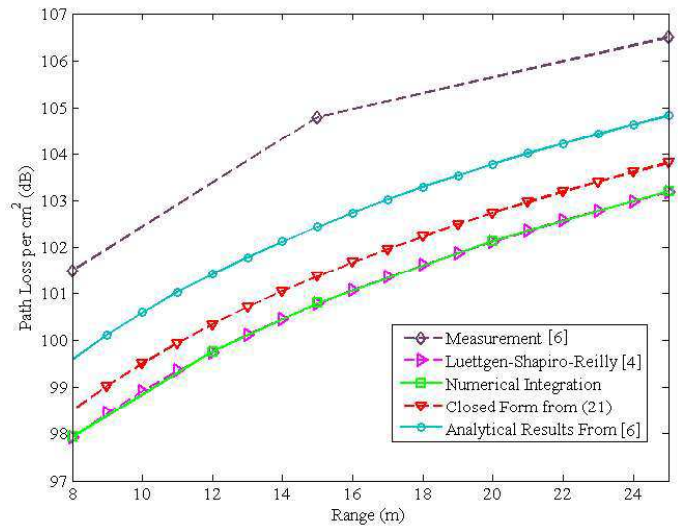

Figure 6. Path loss vs. the range for $\theta_1 = 80°$, $\theta_2 = 60°$, $\phi_1 = 10°$ and $\phi_2 = 30°$, using different methods.

Fig. 7 shows the path losses for geometry $\theta_1 = 20°$, $\theta_2 = 30°$, $\phi_1 = 17°$ and $\phi_2 = 30°$. The measurement and Monte Carlo results are taken from Drost et al.[3] In Fig. 8 the path losses from different methods are plotted for $\theta_1 = 40°$, $\theta_2 = 20°$, $\phi_1 = 10°$, $\phi_2 = 30°$, using the same scattering and extinction coefficient as for Fig. 7. The path loss results from numerical integration and the Luettgen-Shapiro-Reilly method matches exactly. The path loss using the Monte Carlo simulation technique, though considering multiple scattering, is remarkably close to these two plots. The approximated link loss from the closed-form expression given in (21) also has a reasonable accuracy. The measurement and Monte Carlo path losses are plotted using the results in Ding et al.[2]

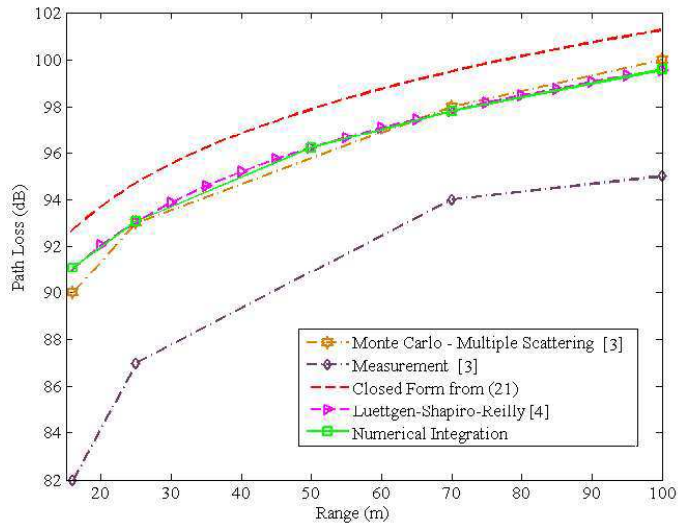

Figure 7. Path loss vs. the range for $\theta_1 = 20°$, $\theta_2 = 30°$, $\phi_1 = 17°$ and $\phi_2 = 30°$, using different methods.

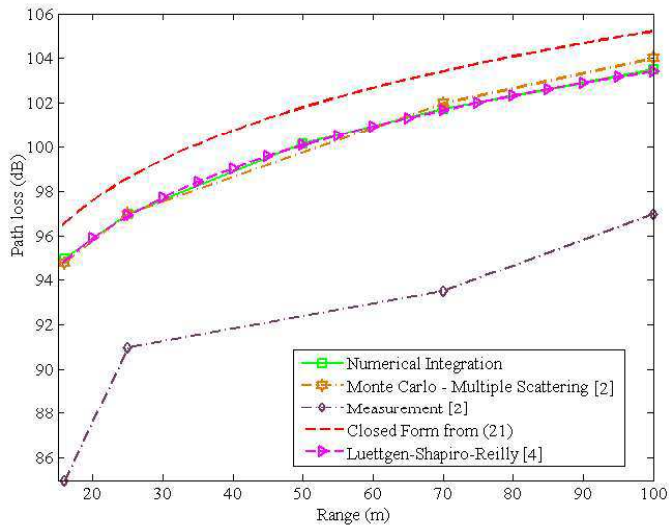

Figure 8. Path loss vs. the range for $\theta_1 = 40°$, $\theta_2 = 20°$, $\phi_1 = 17°$ and $\phi_2 = 30°$, using different methods.

## 5. CONCLUSION

In this paper simulation and analytical modeling methods are proposed for the NLOS UV system. The accuracy and computational complexity of our numerical integration method compare favorably with that of the Monte Carlo simulation, while maintaining all the flexibility. The impulse response and path losses from the proposed approaches are compared to several previously presented techniques. The results obtained using the numerical integration are shown to be in good agreement with the Luettgen-Shapiro-Reilly results for different system geometries. Our approximate closed-form expression is also shown to be more accurate compared to the previously calculated closed-form approximations. In future work we plan to use these techniques to estimate the performance of NLOS UV systems with more complex geometries.

## 6. ACKNOWLEDGMENT

This research was funded by the National Science Foundation (NSF) under grant number ECCS-0901682.